\documentclass[]{openjournal}

\usepackage{latexsym}
\usepackage{graphicx}
\usepackage{amssymb}
\usepackage{longtable}
\usepackage{epsf}
\usepackage{amsmath}
\usepackage{graphicx}
\graphicspath{ {figures/}}
\usepackage{gensymb}
\DeclareMathOperator{\sech}{sech}
\usepackage{hyperref}
\usepackage{cleveref}

\slugcomment{Draft Version \today}
\shorttitle{Bayesian Approach to MW Structure}
\shortauthors{Dobbie P. and Warren S. J.}

\begin{document}

\title{A Bayesian Approach to the Vertical Structure of \\ the Disk of the Milky Way}

\author{Dobbie P.$^{1,2}$ and Warren S. J.$^2$}
\affil{$^1$Sydney, NSW, Australia. \textit{phillip.dobbie@gmail.com}}
\affil{$^2$Astrophysics Group, Blackett Laboratory, Imperial College London, London SW7 2AZ, UK}

\maketitle

This work investigates the vertical profile of the stars in the disk of the Milky Way. The models investigated are of the form $\sech^{2/n}(nz/(2H))$ where, setting $\alpha = 2/n$, the three functions of the sequence $\alpha = 0,1,2$ correspond to exponential, $\sech$, $\sech^2$ functions. We consider symmetric models and asymmetric models, above and below the plane. The study uses the large sample of K and M stars of \cite{1Ferguson} and applies the methods of Bayesian model comparison to discriminate between the 6 models. Two inconsistencies in \cite{1Ferguson}, concerning the vertical height cut and the model continuity across the plane, are noted and addressed. We find that (1) in the Milky Way the symmetric disc models are decisively ruled out, with northern thin disc scale heights $\sim25\%$ larger than southern, (2) there is moderate evidence for the exponential and $\sech$ models over the $\sech^2$ model, though a sample extending further into the Galactic mid-plane is needed to strengthen this result, (3) the photometric distances used by \cite{1Ferguson} underestimate the GAIA distances by a factor of roughly 1.16, and (4) the increase of scale height with Galactic latitude observed by \cite{1Ferguson} is due to incorrect cuts to the data.

\vspace{1cm}

\twocolumngrid

\section{Introduction}
\label{sec:intro}

The vertical profile of the density of stars in the disk of the
Milky Way has been the subject of numerous studies. Over a limited
range of heights $z$ from the plane, the density variation may be
modelled as exponential $e^{-z/H}$, where $H$ is a scale height. Close
to the plane the density profile is assumed to soften, and a
functional form $\sech^2(z/(2H))$ is commonly employed. The $\sech^2$
function asymptotes to the exponential function for $z\gg H$. Beyond a
few scale heights there is an excess in the tail, requiring a second
population, of larger scale height. The two populations are the thin
and the thick disk \citep{1GilmoreReid}. Fitting two
populations provides a useful approximation to the structure of the disk, but this is itself a simplification \citep[e.g.][]{BovyRix2012, Xiang2018}.

The $\sech^2$ and exponential functions are specific examples of
a family of models of the form $\sech^{2/n}(nz/(2H))$ \citep{Kruit1988},
where $n=1$ is $\sech^2$ and $n=\infty$ is exponential. In the
following we use instead $\alpha=2/n$, and consider the three functions
of the sequence $\alpha=0,1,2$, i.e. the 
exponential, $\sech$, $\sech^2$ functions. This is a sequence of successive
flattening in the central plane. Matching $\sech^{\alpha}$ to
the exponential at large distances, the central-plane density is lower 
by the factor $2^{\alpha}$. In summary, \cref{eqn:sech2,eqn:sech1,eqn:exp} are
possible parameterisations of the vertical variation of the stellar
density as a function of vertical distance from the plane of the Milky
Way:

\begin{flalign}
n(z) = n_0\left[e^{-\frac{\lvert z+z_\odot\rvert}{H_1}}+f e^{-\frac{\lvert z+z_\odot\rvert}{H_2}}\right] &&
\label{eqn:exp}
\end{flalign}
\begin{flalign}
n(z) = n_0\left[\sech\left(\frac{z+z_\odot}{H_1}\right)+f\sech\left(\frac{z+z_\odot}{H_2}\right)\right] &&
\label{eqn:sech1}
\end{flalign}
\begin{flalign}
n(z) = n_0\left[\sech^2\left(\frac{z+z_\odot}{2H_1}\right)+f\sech^2\left(\frac{z+z_\odot}{2H_2}\right)\right] &&
\label{eqn:sech2}
\end{flalign}

The parameter $z$ is now the height from the Sun and $z_\odot$ is the height
of the Sun above the plane. The parameters $H_1$ and $H_2$ are the scale heights of the thin disk and the thick disk respectively. The parameter $f$ quantifies the density of the
thick disk in the plane, relative to the density of the thin disk. The
density in the midplane is $n_0(1+f)$. We have used the same variables as \cite{1Ferguson}, and our equation (3) is identical to their equation (7).

The parameter $\alpha$ is of theoretical interest. The case $\alpha=2$ corresponds
to the equilibrium distribution of a uniform planar isothermal population of
stars \citep{0Spitzer}, and is therefore a useful benchmark. The actual
value of $\alpha$ is affected by the potential of the thinner gaseous disk, and is
predicted to be smaller than the isothermal value at the
solar radius \citep{Banerjee,Sarkar2019}.

There have been several measurements of $\alpha$ in external edge-on spiral galaxies. These are best undertaken in the $K$ band to minimise the effects of extinction. The most detailed study of this topic is the analysis of \citet{deGrijs}. For a sample of 24 galaxies they measure a distribution of values of $\alpha=0.5\pm0.2$ (corrected for seeing and extinction), and argue that the true value is even lower, due to a bias because the galaxies are not perfectly edge on.

Most surveys used to analyse the structure of the Milky Way disk
sample a conical volume centred on the Sun. For a conical geometry the volume surveyed
near the Galactic plane is not well suited to establishing the density (and
therefore the flattening) accurately there. Consequently there have
been very few attempts to measure $\alpha$ for the Milky Way, and many
authors simply assume $\alpha=2$. There are no Milky Way studies for
which the measurement of $\alpha$ was a primary
aim. Among the studies of this question, \citet{Hammersley1999} state that ``Analysis of one relatively
isolated cut through an arm near longitude 65 degrees categorically
precludes any possibility of a $\sech^2$ stellar density function for
the disc.'' In contrast, using GAIA DR1, \citet{Bovy2017} states that
for A to K stars ``All vertical profiles are well represented by
$\sech^2$ profiles'', but he does not test other values of $\alpha$. In addition, later than spectral type F the fits are not compelling. Studies with GAIA have the advantage that it is possible to select complete samples within a cylinder, rather than a cone, thereby better sampling the density close to the plane (provided extinction is correctly accounted for, of course).

The analysis of main-sequence turnoff and subgiant stars by
\citet{Xiang2018}, divided into bins of different age, provides the most detailed picture of the vertical structure of the disk in the solar neighbourhood (their
Fig. 18). The measurement of $\alpha$ is a by-product of their study of
the midplane stellar mass density. Their measurements show that
$\alpha$ decreases with age, and for the whole sample (all ages
combined) $\alpha\sim 0$, in disagreement with \citet{Bovy2017}. The selection functions for the studies of \citet{Bovy2017} and \citet{Xiang2018} are both highly complex, especially for the latter,  This means it is impossible to pinpoint where the discrepancy between the results lies without reanalysing the data. Neither survey was aimed specifically at measuring $\alpha$, and a better solution is to undertake a new study tailored specifically to this question.

A recent development in the measurement of the structure of the Milky Way disk has been the recognition that the stars are not in statistical equilibrium. \citet{widrow2012assym} were the first to present evidence for a wave-like north-south asymmetry in the vertical number counts. This was subsequently confirmed in the more detailed studies
of \citet{yanny2013assym}, \citet{1Ferguson}, and \citet{Bennett2019}. The analysis of \citet{1Ferguson} uses a very large sample of several million K and M stars, with photometric parallaxes, in matching footprints above and below the Galactic plane. This allows them to compare measurements of the scale heights above and below the plane, to quantify the asymmetry. They also investigated the shape of the number density distribution for different latitude ranges with a fixed range in longitude and found the scale heights to be sensitive to the selected latitude window. They invoked the different metallicities of the thin and thick disks (leading to errors in the photometric parallaxes) as a possible explanation for this effect.

The starting point for the current study was our interest in exploring whether the large sample of \citet{1Ferguson} might provide an improved measurement of $\alpha$ for the Milky Way \footnote{the measurement of the other parameters, including the scale heights $H_1$ and $H_2$, is not a primary motivation of the current paper; see \citet{1Ferguson} and references therein for a summary of recent measurements.}, and the possibility that the application of Bayesian model comparison techniques \citep[e.g.][]{Trotta08}, which have been little used in this field, might lead to new insights. Early on we identified two inconsistencies in the analysis presented by \citet{1Ferguson}, and we explore the consequences here. The first of these is to do with the vertical height cut used in their analysis; this is explained in section \ref{sec:datasamples}. The second inconsistency is that in considering asymmetric models, in fitting independent scale heights both above and below the Galactic plane, there was no requirement imposed that the densities of the two functions match at the mid-plane. In our own analysis we impose continuity of density through the plane. In the original paper identifying a north-south asymmetry  \citet{widrow2012assym} plotted residuals compared to a smooth symmetric model. Here instead we quantify the asymmetry by comparing scale heights between the North and South Galactic hemispheres i.e. in this paper the word asymmetry equates to a difference in scale heights.

In the current paper we fit six different parametric models for the vertical density distribution to data from eight of the matched fields from \citet{1Ferguson}, correcting for the inconsistencies referred to above. We employ three values of $\alpha=0, 1, 2$, \cref{eqn:sech2,eqn:sech1,eqn:exp}, and we compare symmetric models (the same scale heights above and below the plane) against asymmetric models (different scale heights above and below the plane). We use a Bayesian approach for parameter estimation and model comparison. The aims are to determine whether an asymmetric model is required by the data, and to find if any values of $\alpha$ are ruled out. In Section \ref{sec:methods} we describe the data set used, and the Bayesian formalism. In Section \ref{sec:results} we present the results in terms of the Bayesian evidence for the different models. Section \ref{sec:summary} summarises.

\section{Methods}
\label{sec:methods}

\subsection{Data Selection}
This study uses the sample of K and M stars selected by \citet{1Ferguson} from the ninth data release of the Sloan Digital Sky Survey (SDSS). The complete dataset was kindly supplied to us by Sarah Gardner. Their sample selection built on the earlier works of \citet{widrow2012assym} and \citet{yanny2013assym}. The selection was limited to fields above $30\degree$ in absolute Galactic latitude. The SDSS $gri$ photometry was corrected for Galactic absorption, and limited to  $15.0<r_0<21.5$, and  $1.8<(g-i)_0<2.4$, where the subscript denotes values corrected for absorption. The distances to each star were computed using the  photometric parallax relation provided by \citet{ivezic2008}, which relates absolute magnitude in the $r$ band $M_r$ to the $(g-i)_0$ colour, and includes a metallicity term. A fixed metallicity [Fe/H]$=-0.3$ was assumed. These distances were then refined using an additional small colour term determined by \citet{yanny2013assym}. 

\begin{table}[t]
    \centering
    \begin{tabular}{||c c c c c||} 
         \hline
         Field & $l_1$ & $l_2$ & $b_1$ & $b_2$ \\ [0.5ex] 
         \hline\hline
         1 & 55.0 & 60.0 & 30.0 & 55.1  \\ 
         \hline
         2  & 60.0 & 65.0 & 32.8 & 59.3\\
         \hline
         3 & 70.0 & 75.0 & 37.2 & 65.3 \\
         \hline
         4 & 95.0 & 105.0 & 48.6 & 72.8 \\
         \hline
         5 & 105.0 & 120.0 & 49.0 & 73.4 \\  
         \hline
         6  & 120.0 & 135.0 & 48.7 & 73.6\\
         \hline
         7 & 135.0 & 150.0 & 46.7 & 71.7 \\
         \hline
         8 & 150.0 & 165.0 & 51.0 & 68.5 \\ 
         \hline
        \end{tabular}
    \caption{The Galactic longitude (l) and Galactic latitude (b) limits for the eight paired fields investigated; $b_1<b<b_2$ and $l_1<l<l_2$. Each field has a paired field at negative $b$. }
    \label{tab:limits}
\end{table}

\subsection{The Data Samples}
\label{sec:datasamples}

\begin{figure*}[ht]
\begin{center}
\includegraphics[scale=0.19]{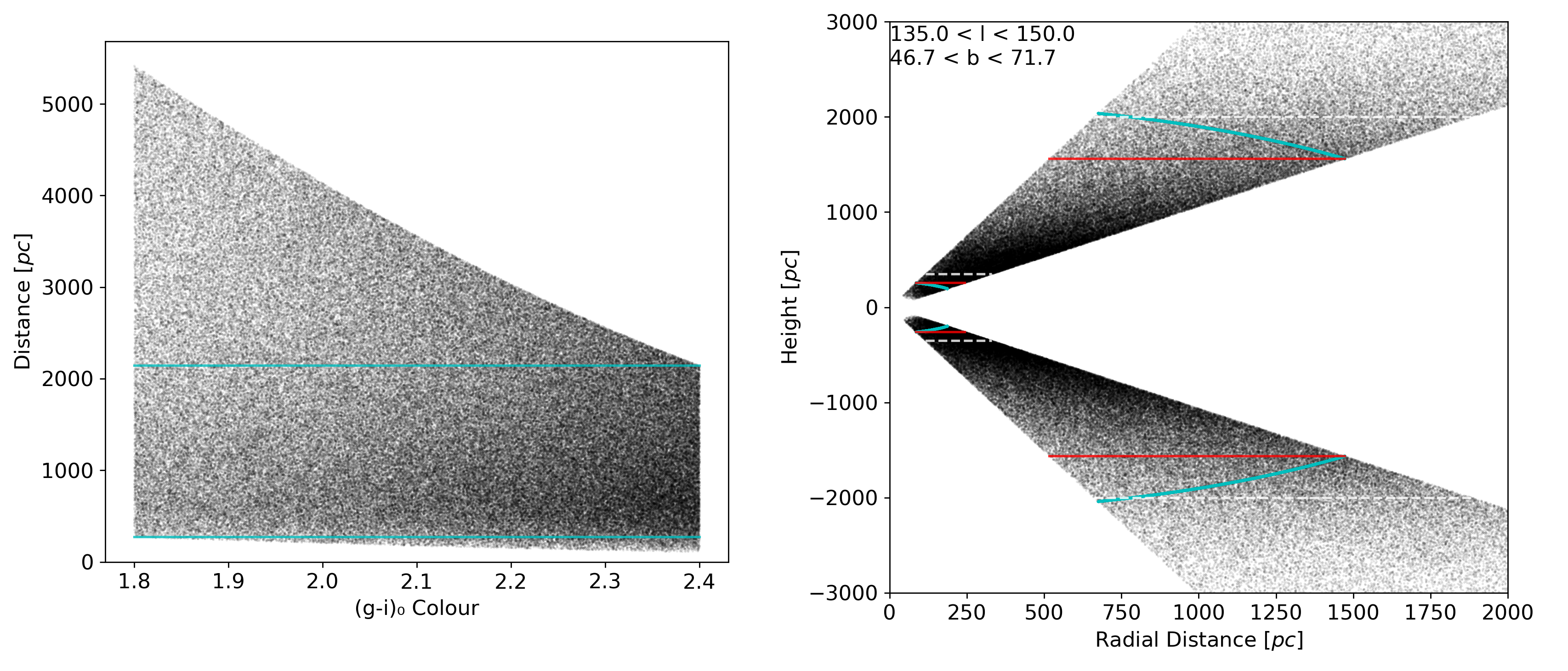}
\caption[Complete Sample]{{\em LHS:} Plot of distance against colour for the subsample $135.0<l<150.0$ and $46.7<|b|<71.7$. The sample's lower and upper distance limits correspond to the apparent magnitude limits $r_0=15.0$ and $r_0=21.5$ respectively. The sample is complete for all colours only over the distance range marked by the blue lines. {\em RHS:} The sample plotted in cylindrical coordinates $z$ against cylindrical radius $r=\sqrt{x^2+y^2}$. The blue lines are again the distance limits shown in the left-hand plot. The dashed white lines mark the $z$ limits used by \citet{1Ferguson}. The red lines mark the $z$ limits used in the current paper. \label{fig:heightcut}
}
\end{center}
\end{figure*}

The \citet{1Ferguson} dataset comprises several subsamples. Each subsample covers an area matched above and below the Galactic plane, and is defined by limits in Galactic longitude $l$ and absolute Galactic latitude $\lvert b\rvert$ (see their Fig. 3). There is one large matched data set, which we do not use, and a set of 14 smaller (individually) matched datasets. From these 14 we selected those samples which cover a wide range in both $l$ and $b$ and also contain a greater number of stars. The Bayesian method described in section 2.3 is best suited to larger data samples, and the nature of the subsample cutting done by \citet{1Ferguson} leaves some subsamples with far fewer stars within their volume. Of the 14 subsamples 8 were chosen as suitable to the analysis here and are listed in Table \ref{tab:limits}.

\citeauthor{1Ferguson} limit their analysis to the height range $0.35\leq|z|\leq2.0$\,kpc, stating that the sample is complete in each field over the individual $l$ and $b$ ranges within the volume defined by these limits, and over the full range of colours $1.8<(g-i)_0<2.4$. The justification that the sample is complete for all colours out to the upper limit comes from \citet{yanny2013assym} (based on their Fig. 16). This argument is incorrect, as we now show. 

In any direction on the sky there are lower and upper limiting distances, corresponding to the apparent magnitude range, that depend on colour. These limits are visible in the left-hand panel of Fig. \ref{fig:heightcut}, where the stars of one particular field are plotted. The sample is complete over all colours only within the distance range marked by the blue lines i.e. between $d_{min}$ for the bluest stars, hereafter $d_{min(blue)}$, and $d_{max}$ for the reddest stars, hereafter $d_{max(red)}$. The same distance limits apply to all fields. These distance limits are shown in the polar plane, plotted in the right-hand panel. The $z$ limits adopted by \citeauthor{1Ferguson} are shown by the dashed white lines.

Here it can be seen that horizontal slices are not complete over the full $b$ range of the field for slices above the upper red line, which is lower than $z=2$\,kpc. The height of the red line, the correct upper completeness limit, is given by the expression $z_{lim(u)}=d_{max(red)}\sin(b_{min})$. 
The incompleteness introduced by adopting the wrong upper limit depends on the latitude range of the field, becoming more severe at lower latitudes. This might lead to visible trends with $b$, that are not real, of parameters of the measured Galactic structure. We return to this point in Section \ref{sec:trends}.

There is a corresponding lower completeness limit, also marked by a red line, given by the expression $z_{lim(l)}=d_{min(blue)}\sin(b_{max})$. \citet{1Ferguson} adopt a lower height limit of 0.35\,kpc. ``due to brightness saturation effects''. However SDSS images are not saturated at $r=15$, so we use instead $z_{lim(l)}$ as given above, which is smaller than 0.35\,kpc in all fields.
This is useful because probing closer to the Galactic plane improves the measurement of $\alpha$. 

The stellar number counts are divided into 50 bins, equally spaced in $z$ between $z_{lim(l)}$ and $z_{lim(u)}$. The volume of a slice is
\begin{equation}
V(z) = \frac{1}{2}\delta(l_2-l_1)z^2\left(\frac{1}{\sin^2b_1}-\frac{1}{\sin^2b_2}\right)
\end{equation}

where $\delta$ is the vertical thickness of the bin, $z$ is the height of the middle point of the bin, and the ranges of longitude and latitude are $l_1<l<l_2$ and $b_1<|b|<b_2$. This is also the formula used by \citet{1Ferguson}. It would, of course, be possible to compute the space density at heights greater than $z_{lim(u)}$ by selecting only sources within $d_{max(red)}$, with appropriate modification of the expression for the volume, but we chose not to, for simplicity.

The space density of stars in the disk varies not only in the vertical direction but also as a function of Galactic radius. In their analysis \citet{1Ferguson} found that allowing for the variation in density with Galactic radius did not affect the extracted parameters in a significant way. For this reason they chose to present their results taking no account of this variation. We have elected to follow the same approach in order that we may compare our results with theirs directly.

\subsection{Bayesian methodology}

For parameter estimation and model comparison we use the PyMultiNest
\citep{buchner2014} implementation of the MultiNest nested sampling algorithm \citep{Feroz09}.

\subsubsection{Parameter Estimation}
\label{sec:parameter}
For observed data $D$, background information $I$, and a model specified by a vector of parameters $\theta$, Bayes' theorem
\begin{equation}
\mathrm{prob}(\theta|D,I) = \frac{\mathrm{prob}(D|\theta,I)\mathrm{prob}(\theta|I)}{\mathrm{prob}(D|I)}
\label{eqn:bayes}
\end{equation}
formulates the task of parameter estimation. The term on the left, the posterior, is the result sought, the probability of the parameters given the data. On the right, the first term in the numerator is the likelihood, which is the probability of the data given the parameters, and the second term is the prior for the parameters. The denominator is the Bayesian evidence, explained in Section \ref{sec:parameter}, which is a normalisation constant that is not needed for parameter estimation.

\begin{figure*}[t]
\begin{center}
\includegraphics[scale=0.24]{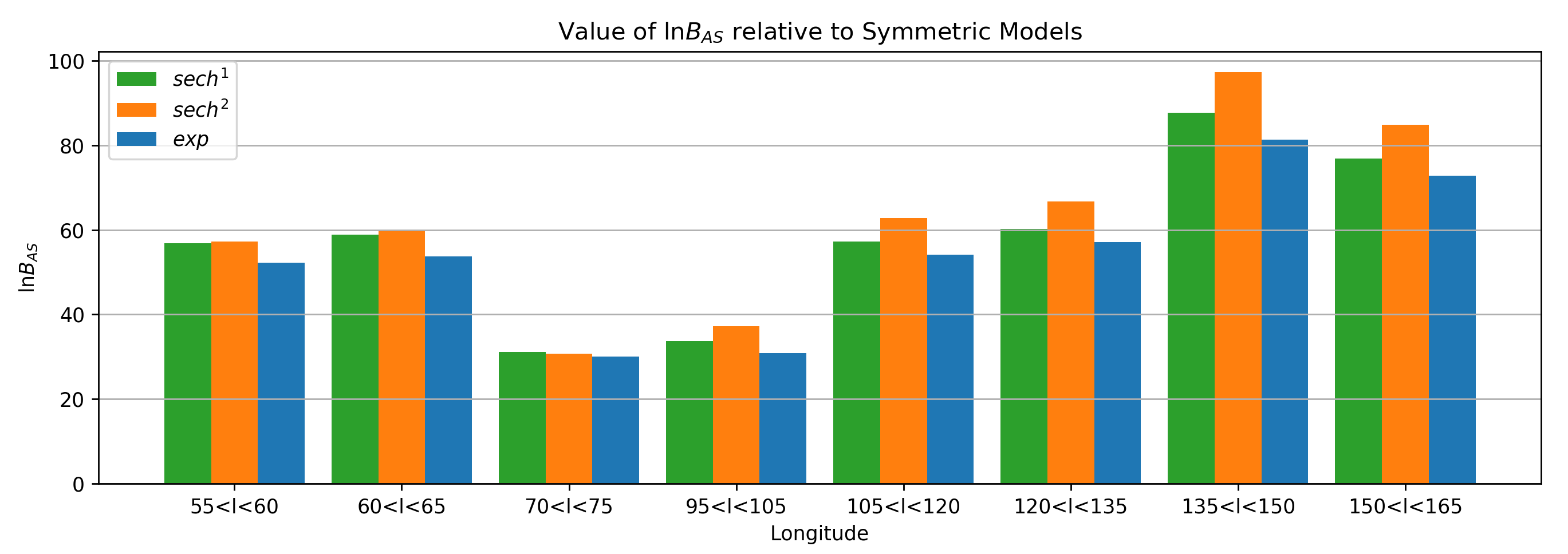}
\caption[Evidence Relative]{The ln$B_{AS}$ values of the asymmetric models shown relative to the symmetric models for the same field.
}
\label{fig:deltae1}
\end{center}
\end{figure*}

Symmetric disk models, for each of the three model profiles, \cref{eqn:sech2,eqn:sech1,eqn:exp}, are specified by five free parameters; $n_0$, $H_1$, $H_2$, $z_\odot$, and $f$. We adopt broad uniform linear priors over the following ranges 
\begin{align}
0.0 < &\:n_0 < 10^{9} \:\:\mathrm{kpc^{-3}}\nonumber \\
0.0 < &\:H_1 < 2.0 \:\:\mathrm{kpc}\nonumber \\
0.0 < &\:H_2 < 2.0 \:\:\mathrm{kpc}\nonumber \\
0.0 < &\:z_\odot < 0.1 \:\:\mathrm{kpc}\nonumber \\
0.0 < &\:f < 1.0
\end{align}
We discuss the choice of priors where it is relevant at various points below.

For asymmetric models we must specify independent scale heights above (N) and below (S) the plane, $H_{1N}$, $H_{2N}$, $H_{1S}$, $H_{2S}$, meaning there are seven free parameters. The same broad uniform prior is used for all scale heights. In contrast to \citet{1Ferguson}, the density models above and below the plane share the same values of $n_0$ and $f$, ensuring that the density distribution is continuous through the disk mid-plane.

For any model, the likelihood is given by
\begin{equation}
\mathrm{prob}(D|\theta,I) = \mathcal{L} = \prod_i\frac{1}{\sqrt{2\pi m_i}}e^{-\frac{(n_i-m_i)^2}{2m_i}}
\label{eqn:likelihood}
\end{equation}
Here $n_i$ is the number of stars in slice $i$ and $m_i$ is the number of stars predicted by the model, and we have approximated the Poisson distribution by the Gaussian distribution since the number of stars in each slice  is large. In practise the sampler uses the negative log likelihood (excluding a constant term)
\begin{equation}
-\ln\mathcal{L} = \sum\frac{(n_i - m_i)^2}{2m_i} + \frac{1}{2}\sum\ln m_i
\label{eqn:loglike}
\end{equation}
Because the likelihood functions for these datasets are very narrow, the choice of priors does not significantly impact the estimation of parameters, since over a narrow range the prior may be treated as
constant and so does not skew the likelihood.

\begin{figure*}[t]
\begin{center}
\includegraphics[scale=0.53]{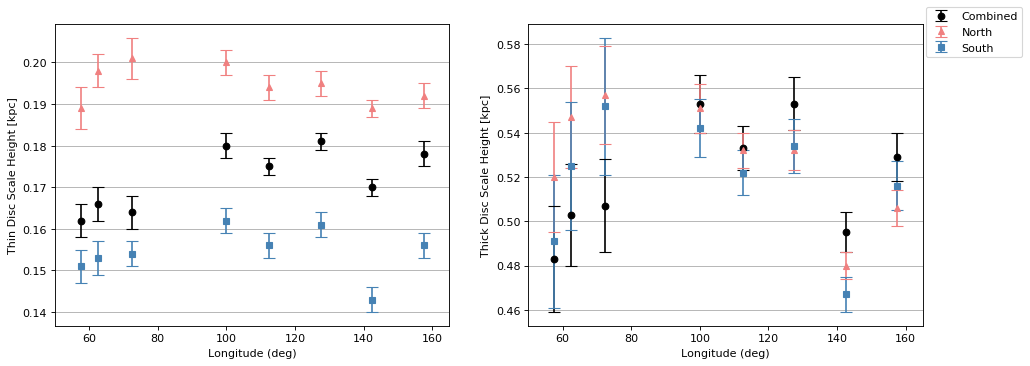}
\caption[Change in Scale Heights]{Scale height comparison for the $\sech^2$ models. Thin disc scale heights shown in figure (a) and thick disc scale heights in figure (b). Circles represent the symmetric model, while triangles and squares are the North and South hemispheres of the asymmetric model respectively.
}
\label{fig:scaleheight}
\end{center}
\end{figure*}

\subsubsection{Model Comparison}
\label{sec:modelcomparison}

Model comparison requires the calculation of the Bayesian evidence for each model, the denominator on the RHS of equation \ref{eqn:bayes}. The Bayesian evidence is the integral of the numerator over the parameter space i.e. 
\begin{equation}
\mathrm{prob}(D|M_i,I) = \int \mathrm{prob}(D|\theta_i,M_i,I)\mathrm{prob}(\theta_i|M_i,I)d\theta_i 
\label{eqn:evidence}
\end{equation}
where $M_i$ specifies the model. Application of Bayes' theorem then leads to the following expression for the ratio of posteriors for two different models, $M_1$, $M_2$,
\begin{equation}
\frac{\mathrm{prob}(M_1|D,I)}{\mathrm{prob}(M_2|D,I)} = \frac{\mathrm{prob}(D|M_1,I)\mathrm{prob}(M_1|I)}{\mathrm{prob}(D|M_2,I)\mathrm{prob}(M_2|I)}
\label{eqn:bayes2}
\end{equation}
Here $\mathrm{prob}(M_i|I)$ is the prior for a model. We assume equal priors in all cases, in which case the problem reduces to computing the Bayes factor, given by
\begin{equation}
B_{12}=\frac{\mathrm{prob}(D|M_1,I)}{\mathrm{prob}(D|M_2,I)}
\label{eqn:factor}
\end{equation}
\citet{Trotta08} tabulates threshold values for considering a model to be superior, with $|\ln B_{12}|>2.5$ listed as providing moderate evidence, and $|\ln B_{12}|>5$ listed as providing strong evidence.

We argued above that the choice of priors does not impact the estimation of the parameters for the datasets used here. In comparing models, though, the priors are important. Although models with larger numbers of parameters will usually provide a better fit to the data, this is balanced in the calculation of the Bayesian evidence by the increased volume of parameter space over which the integral is performed. In comparing the symmetric models to the asymmetric models, the width of the priors for the extra scale heights then become important. This is because, for uniform priors over a width $H_{upper}-H_{lower}$, the probability density of the prior is inversely proportional to the width. This point must be borne in mind when considering the results, in the next section.

\section{Results and discussion}
\label{sec:results}

In this study we are more interested in the comparison of different models than the values of the measured parameters. Three cases of $\alpha$ are considered, $\alpha=0,1,2$, and we fit both symmetric and asymmetric models. These six models are each fit to the data from the  eight fields listed in Table \ref{tab:limits}, providing 48 sets of results.

\begin{figure*}[t]
\begin{center}
\includegraphics[scale=0.25]{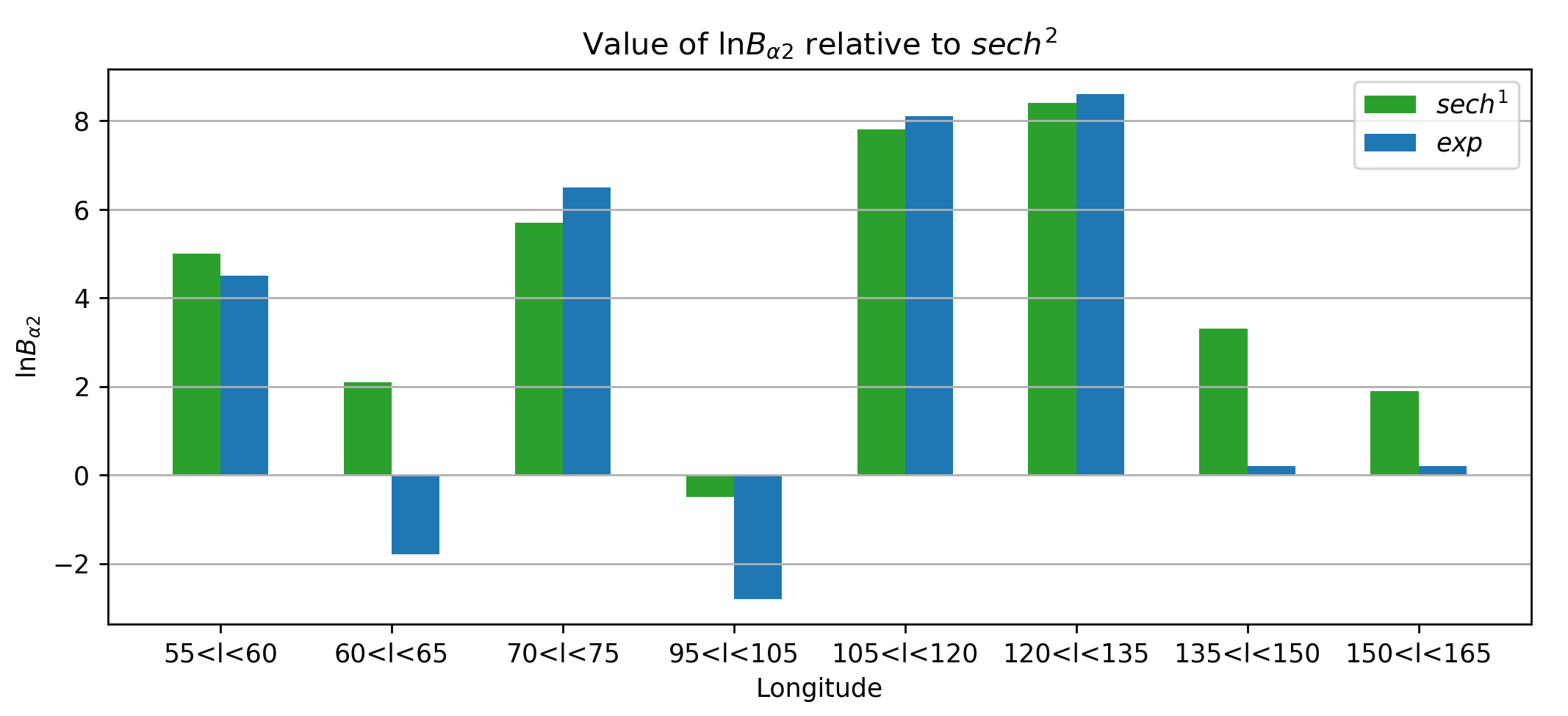}
\caption[Evidence Relative]{The ln$B_{\alpha2}$ values for exponential ($\alpha=0$) and $\sech$ ($\alpha=1$) models relative to $\sech^{2}$. The greater the value of $|\ln B_{\alpha2}|$, the better suited the model is to the data relative to $\sech^{2}$. 
}
\label{fig:deltae2}
\end{center}
\end{figure*}

\begin{figure*}[t]
\begin{center}
\includegraphics[scale=0.32]{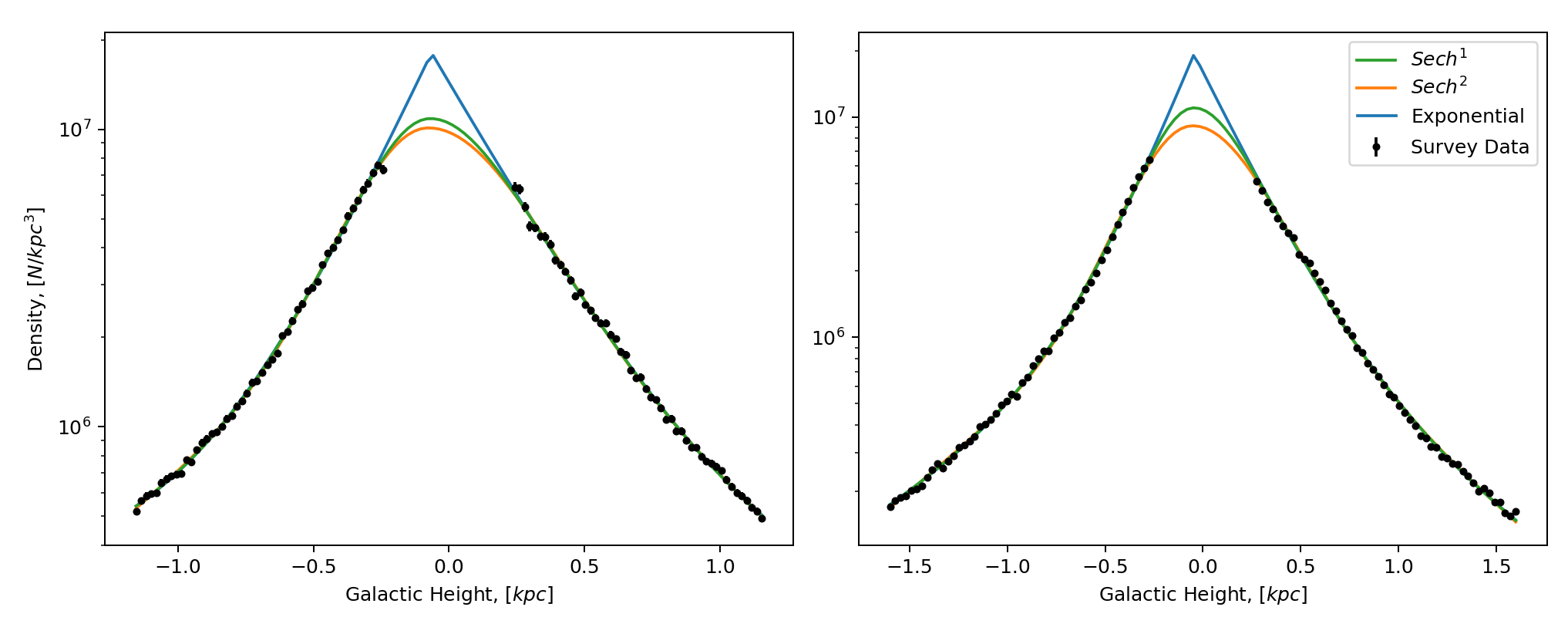}
\caption[Master Plot]{The three asymmetric model types plotted together for the data slices of $60<l<65$ (a) and $120<l<135$ (b) in log-space. 
}
\label{fig:masterplot}
\end{center}
\end{figure*}

\subsection{Comparison of symmetric and asymmetric models}
We consider firstly the evidence for asymmetry in the density distribution, above and below the plane. We have measured the Bayes factor comparing the asymmetric model (A) to the symmetric model (S), $B_{AS}$, for each value of $\alpha$, in each field, providing 24 values of $B_{AS}$. The values of $\ln(B_{AS})$ are plotted in Fig. \ref{fig:deltae1}. They range from 30 to 90. This is decisive evidence for the asymmetric model over the symmetric model. This is despite the rather broad priors used for all the scale heights $0<H<2$\,kpc, which act to penalise the model with more parameters. 

The large values of $B_{AS}$ are not particularly surprising, given that asymmetry in the density distribution is clearly visible in plots of the binned space density above and below the plane, e.g. figures in \citet{1Ferguson}. Nevertheless it is satisfying to see this so strongly confirmed by the Bayes factor analysis, and it suggests that model comparison could identify more subtle effects that might not be so readily apparent in the binned data.

The measured values of the respective scale heights for the thin and thick disks, for the symmetric and asymmetric models are plotted in Fig. \ref{fig:scaleheight}, specifically for the $\alpha=2$ models. For the thick disk (RHS) the scale heights above and below the plane are in good agreement. For the thin disk the scale heights above the plane are about $25\%$ larger than the scale heights below the plane. The average North and South thin disk scale heights are $0.195\pm 0.004$\,kpc and $0.155\pm 0.006$\,kpc respectively (the quoted uncertainties are the scatter in the measurements) and the measured values of the thin disk scale height in the Northern hemisphere range between 18\% and 34\% greater than those in the Southern hemisphere. The actual values measured are discussed below.

\subsection{Comparison of different values of $\alpha$}

Because the asymmetric models are decisively preferred, in comparing the fits for different values of $\alpha$ we consider the asymmetric models only. In most cases the $\alpha=2$ model i.e. $\sech^2$ provides the worst fit. In Fig. \ref{fig:deltae2} we plot the eight measured values of $\ln B_{02}$ and of $\ln B_{12}$  i.e. the values of the Bayes factors  for the  exponential and $\sech$ models relative to the $\sech^2$ model. The average values of $\ln B_{02}$ and of $\ln B_{12}$ are 2.9 and 4.2 respectively. This is moderate evidence against the $\sech^2$ model.

Examples of fits of the three different asymmetric models for two of the fields are provided in Fig. \ref{fig:masterplot}. 
In each field the three models appear very similar over the height ranges which are fit, i.e. where there is data, but the models are strikingly different at heights $<~300$\,pc i.e. close to the plane where there is no data. In this respect it is impressive that the model comparison provides moderate evidence against the $\sech^2$ model. It appears likely that applying the same model comparison method to a survey with data closer to the plane could definitively distinguish between the three different values of $\alpha$. 

The measured asymmetric scale heights for a particular model are similar between different fields, as shown in Fig. \ref{fig:scaleheight}. Typical values of the scale height for the thin disk are $\sim 200$\,pc for the $\sech$ model, with slightly smaller (larger) values measured for the $\sech^2$ (exponential) models. We measure smaller scale heights than \citet{1Ferguson}. This statement applies even when we attempted to reproduce their results using the same cuts and fitting methods that they used, so it is not a consequence of these factors, and we have not found an explanation. Our measured scaleheight for the thin disk $\sim 200$\,pc is noticeably smaller than the fiducial value of 300\,pc \citep[e.g.][]{1Bochanski,1GilmoreReid,1Chang,1Juric}. We have found that part of the discrepancy lies with the photometric parallaxes used by \citet{1Ferguson}. To check their distances we matched a subsample to the GAIA DR2 data release \citep{Gaia2}. We started with all the sources in the N half of the field $48.7<b<73.6$, $120<l<135$, and selected sources matched within 1\,arcsec, with parallax/error$>30$, producing a sample of 5391 sources with accurate geometric parallaxes. We computed the ratio of the photometric distance over the geometric distance $d_{ph}/d_{GAIA}$. The median value of this ratio is 0.86 with a $1\sigma$ scatter of 0.12. This indicates that the photometric distances should be multiplied by a factor 1.16, implying a scale height of $\sim230$\,pc for the $\sech$ model.

A further correction needs to be made for the presence of binaries in the photometric sample. Applying this correction will act to increase the scale height \citep[e.g.][]{Covey09,1Bochanski}. Alternatively we can compare against the scaleheight of $255$\,pc measured by \citet{1Bochanski} which is the measured scaleheight before correction. These two numbers are now in reasonable agreement. In fact the correction for binaries will be somewhat larger for the sample of \citet{1Ferguson} compared to that of \citet{1Bochanski}, because the stars are of earlier spectral type, on average, for which the binary fraction is larger \citep{Duchene13}. This means that the agreement is even better.

\begin{figure}[t]
\begin{center}
\includegraphics[scale=0.5, trim= 1.5cm 6.cm 0.cm 5.cm]{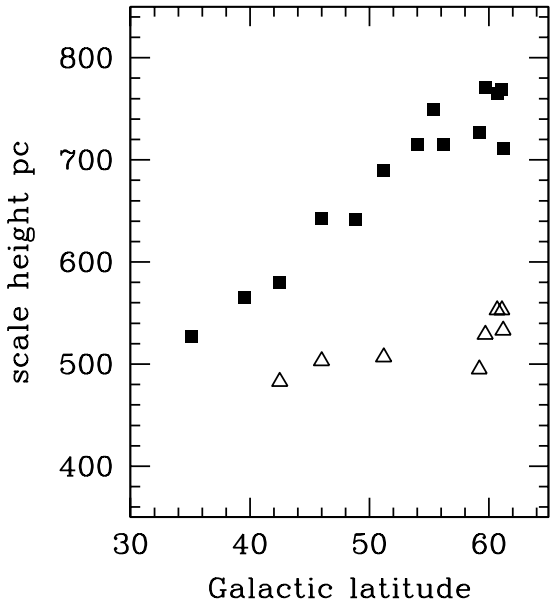}
\caption[]{The scale height of the thick disk as a function of Galactic latitude $b$. Filled squares plot values from \citet{1Ferguson}, and open triangles plot values from this paper. In each case the data are for the symmetric $\sech^2$ model.}
\label{fig:fergthick}
\end{center}
\end{figure}

\subsection{Trends with Galactic latitude}
\label{sec:trends}

\citet{1Ferguson} noted a trend between measured scale height and average Galactic latitude of each field measured, for both the thin and the thick disk. They ascribed this to metallicity trends that affect the distance estimates. In Fig \ref{fig:fergthick} their measured scale heights for the thick disk are plotted, as filled squares, against average $b$ for each field. These data are for the symmetric model and were extracted from their Fig. 7 (14 fields). There is a strong trend with $b$, and a similar trend is seen for their thin disk measurements. Our own measurements (8 fields) for the thick disk are plotted as open triangles, and do not display a strong trend with $b$. The same is true for our thin disk results. This indicates that the strong trend seen in the results of \citet{1Ferguson} is an artifact caused by the incorrect distance cuts they applied, as anticipated in Section \ref{sec:datasamples}.

\section{Summary}
\label{sec:summary}

We have used Bayesian model comparison to investigate the vertical structure of the disk of the Milky Way at the solar radius, using the large homogeneous sample of \citet{1Ferguson}. We corrected for two inconsistencies in the previous analysis by \citet{1Ferguson}, one to do with the completeness limits of the sample, and the other by ensuring that asymmetric density models above and below the plane are continuous through the mid plane. The main findings are:

1. Symmetric models for the vertical profile of the Galactic disk are decisively ruled out, based on the measured Bayes factors. The scale heights of the thin disk in the N are $\sim25\%$ larger than in the S. 

2. There is moderate evidence for the exponential or $\sech$ models, respectively $\alpha=0,1,$ over the $\sech^2$ model, but a sample extending closer to the Galactic mid-plane is needed to strengthen this result.

3. The photometric distances used by \citet{1Ferguson} underestimate the true distance by a factor 1.16 on average.

4. We have presented evidence that the strong trend of scale height
with Galactic latitude $b$ found by \citet{1Ferguson} is due to
incorrect cuts applied to the data.

After completing the calculations presented in this paper a new sample
appeared which is highly suitable for analysis by the methods
developed here \citep{ahmed2019homogeneous}. The analysis of that sample will be presented in a forthcoming paper.

\begin{acknowledgements}
We are grateful to Sarah Gardner who provided the tables of the sample of \citet{1Ferguson}. We thank the anonymous referee for suggestions that improved the presentation.
\end{acknowledgements}

\bibliographystyle{ECA_jasa}
\bibliography{bibliography}

\end{document}